\documentclass[%
reprint,
amsmath,amssymb,  aps, prc, nofootinbib
]{revtex4-1}

\usepackage{graphicx}
\usepackage{dcolumn}
\usepackage{bm}
\usepackage[utf8]{inputenc}
\usepackage[T1]{fontenc}

\begin{document}


\title{Cross-conductivity: novel transport coefficients to constrain the hadronic degrees of freedom of nuclear matter}

\author{ J.-B.~Rose$^{1,2}$, M.~Greif$^{2}$, J.~Hammelmann$^{1,2}$,
J.A.~Fotakis$^{2}$, G.S.~Denicol$^{3}$, H.~Elfner$^{4,2,1}$ and C.~Greiner$^{2}$}

\affiliation{$^1$Frankfurt Institute for Advanced Studies, Ruth-Moufang-Strasse 1, 60438
Frankfurt am Main, Germany}
\affiliation{$^2$Institute for Theoretical Physics, Goethe University,
Max-von-Laue-Strasse 1, 60438 Frankfurt am Main, Germany}
\affiliation{$^3$Instituto de F\'isica, Universidade Federal Fluminense, UFF, Niter\'oi, 24210-346, RJ, Brazil}
\affiliation{$^4$GSI Helmholtzzentrum f\"ur Schwerionenforschung, Planckstr. 1, 64291
Darmstadt, Germany}


\date{\today}

\begin{abstract}
In general, the constituents of the bulk matter produced in heavy-ion collisions carry, besides electric charge, multiple other conserved quantum numbers like baryon number and strangeness. Therefore, an electric field will not only generate an electric current but, at the same time, also currents in baryon number and strangeness. We propose that the impact of the electric field on these conserved currents should be characterized by additional transport coefficients, which we call cross-conductivities. In this paper, we introduce and present a calculation of these cross-conductivities from the Green-Kubo formalism within the transport code SMASH for different chemical compositions of hadron resonance gases. We find that the coefficients underlie an ordering in the active degrees of freedom and that thus the chemical composition of the system plays a crucial role. Further, we argue that in future comparisons of lattice QCD calculations with these findings, one could constrain which degrees of freedom and their corresponding charge properties are relevant for the QCD dynamics of the system.\end{abstract}

\maketitle

The well-known electrical conductivity describes the response of a medium either to an external electric field or an uneven distribution of charge density, and is consequently only sensitive to transport cross sections involving charged particles. In the field of heavy ion physics, the electrical conductivity is used in theoretical predictions of the low-mass dilepton yield \cite{Ding:2016hua,Ghiglieri:2016tvj}; it has also been related to the diffusion of magnetic fields in a medium \cite{Baym:1997gq,Tuchin:2013ie,FernandezFraile:2005ka} and is an important input to magnetohydrodynamics \cite{Roy:2017yvg,Inghirami:2016iru,Roy:2015kma}, allowing for a longer duration of the strong initial magnetic field when non-zero. This has produced many explicit calculations of this conductivity using various formalisms \cite{Greif:2016skc,Greif:2014oia,Puglisi:2014sha,Cassing:2013iz,Steinert:2013fza,Marty:2013ita,Berrehrah:2015vhe,Finazzo:2013efa,Finazzo:2014cna,Rougemont:2017tlu,Ding:2016hua,Amato:2013naa,Aarts:2014nba,Brandt:2015aqk,Qin:2013aaa,Arnold:2000dr,Arnold:2003zc,Mitra:2016zdw,Hattori:2016cnt,Fukushima:2017lvb}.

In this work, we introduce two new transport coefficients that quantify the effect that the electric field must also have on other conserved-charge currents: the baryon-electric conductivity, and the strange-electric conductivity. Together with the previously discussed electric conductivity, we shall refer to this set of coefficients as the ``cross-conductivity''. Individual hadrons that constitute the matter below the phase transition to the quark-gluon plasma carry not only an electrical but sometimes a combination of several conserved charges. Prominent examples include the proton, that is a baryon and has positive electric charge or the kaon which is a strange meson that also carries an electric charge. The same is true for quarks themselves, but in the current work we focus on the hadronic cross-conductivities. Generally, the application of an electric field can produce not only an electric, but also a baryonic or strange current, such that
\begin{align}
\begin{split}
\textbf{j}_Q &= \sigma_{QQ} \textbf{E}, \\
\textbf{j}_B &= \sigma_{QB} \textbf{E},\\
\textbf{j}_S &= \sigma_{QS} \textbf{E}.
\end{split}
\label{constitutive_QBS}
\end{align}
where $\textbf{E}$ is an external electric field and $\textbf{j}_{Q,B,S} $ are respectively the electric, baryonic and strange currents that represent the medium response, when it is subjected to an external electric field or uneven charge distributions. If calculated in systems of various chemical compositions, the differences in the behavior of these transport coefficients can be used to learn more about the active degrees of freedom of hadronic matter. As was pointed out by comparing lattice and hadronic equations of state \cite{Huovinen:2009yb}, partial pressures, especially in the strange baryon sector \cite{Alba:2017mqu}, and baryon-strange and strange correlations, as well as ratios thereof \cite{Bellwied:2019pxh,Bazavov}, our current understanding of how many such active degrees of freedom have to be included in hadronic effective models to fully reproduce the underlying QCD dynamics is limited. More precise constraints independent of experimental measurements are highly desirable. This work, through the calculation of the cross-conductivity in systems of increasing number of degrees of freedom, aims to provide a novel way in which this can be accomplished, notably through a comparison with future lattice QCD calculations.

Additionally, note that in the hadronic phase, some semianalytical calculations of the electrical conductivity in pion gases were made using chiral perturbation theory \cite{FernandezFraile:2005ka,Torres-Rincon:2012sda}, in a pion gas using a sigma model where the effect of including medium-modified interactions was studied \cite{Ghosh:2018kst}, in a sigma model with baryonic and mesonic interactions \cite{Ghosh:2016yvt}, and finally in resonance gas models \cite{Greif:2016skc,Cassing:2013iz,Steinert:2013fza}. Almost none of the previously mentioned calculations include a dynamical treatment of resonances, and thus this work further contributes a solid determination of the electrical conductivity of a realistic hadronic gas (as far as our current understanding of the active degrees of freedom goes) which takes into account these effects. To the knowledge of the authors, no previous calculations currently exist for the baryon-electric and strange-electric cases.

We use the Green-Kubo formalism to express the cross-conductivity in terms of correlation functions, such that
\begin{align}
\sigma_{QQ} &= \frac{V}{T} \int^{\infty}_0 \langle j_Q^x j_Q^x (t') \rangle_0 \: dt', \label{kubo_QQ} \\
\sigma_{QB} &= \frac{V}{T} \int^{\infty}_0 \langle j_B^x j_Q^x (t') \rangle_0 \: dt', \label{kubo_QB} \\
\sigma_{QS} &= \frac{V}{T} \int^{\infty}_0 \langle j_S^x j_Q^x (t') \rangle_0 \: dt', \label{kubo_QS}
\end{align}
with $V$ the volume of the system, $T$ its temperature, and $\langle \ldots \rangle_0$ denoting an averaging over an equilibrium state. Eq. \eqref{kubo_QQ} is the well-known Green-Kubo relation for the electric conductivity, and it is quite straightforward to show the correctness of Eqs. \eqref{kubo_QB} and \eqref{kubo_QS} using derivations which are in all points similar to textbook derivations of the electric conductivity \cite{Zubarev}, by substituting the appropriate currents as shown in Eq. \eqref{constitutive_QBS}.

\begin{figure*}
  \centering
  \includegraphics[width=140mm]{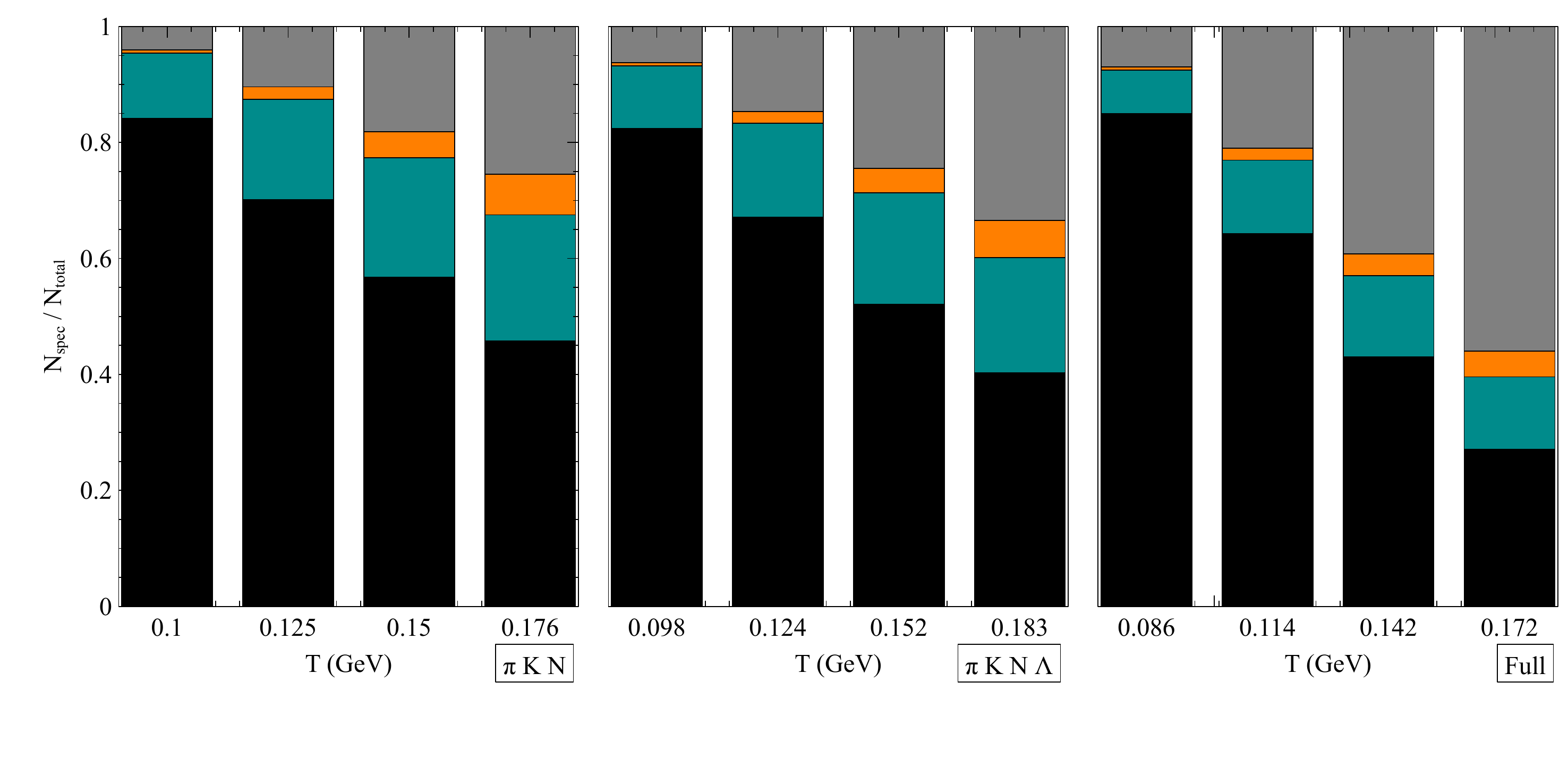}
  \caption{Chemical composition of the $\pi K N$, $\pi K N \Lambda$ and full hadron gases at various temperatures. Pions are depicted in black (first from bottom in every column), kaons in dark cyan (second from bottom), nucleons in orange (third from bottom), and all other species compose the grey section (top).}
  \label{chem_comp}
\end{figure*}

\begin{figure}
  \centering
  \includegraphics[width=76mm]{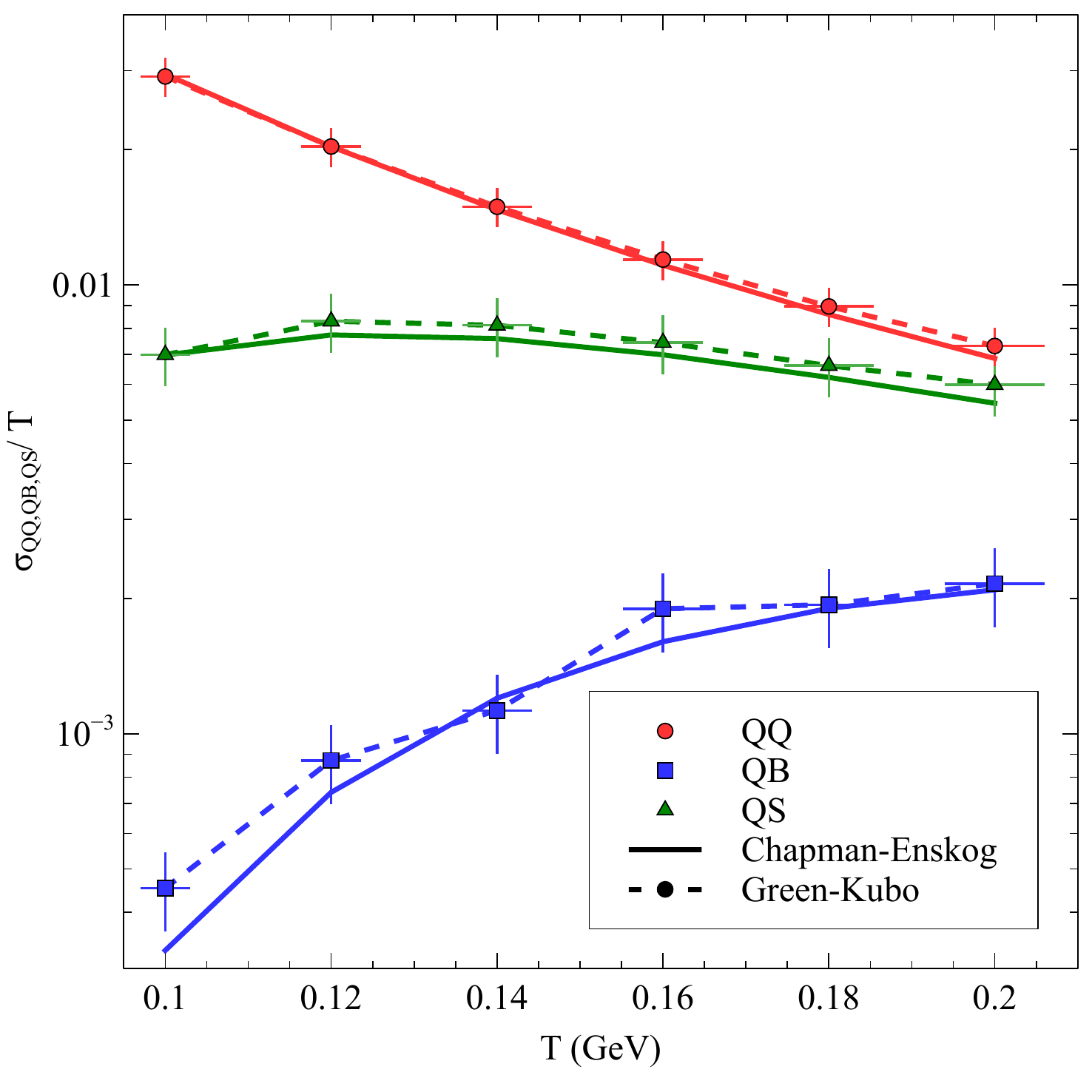}
  \caption{Simple $\pi-K-N$ gas cross-conductivity interacting through a constant cross-section of 30 mb, as computed following a Chapman-Enskog approach~\cite{Hammelmann:2018ath,Greif:2016skc,Greif:2017byw} (dashed lines) and using the Green-Kubo formalism (dotted lines and symbols).}
  \label{calibration_cross}
\end{figure}

In this paper we calculate the transport coefficients defined in Eqs. \eqref{kubo_QQ}-\eqref{kubo_QS} for an interacting hadron gas. This hadron gas is simulated in a box with periodic boundary conditions using the transport code SMASH  \cite{Weil:2016zrk,dmytro_oliinychenko_2019_3485108}, which uses a geometric collision criterion and allows access to the full phase-space history at regular intervals. In the following we will be looking at three different chemical compositions for the hadron gas, all of which are initialized with thermal multiplicities. The first one is a simple mixture of pions, kaons and nucleons, and is the simplest example of a hadron gas containing all three relevant charges (electric, baryonic and strange) for this discussion. We will be investigating both the cases where this gas interacts as hard spheres (dubbed $\pi K N$-hard) and through resonance formation ($\pi K N$) in the following pages. The second one is a mixture of pions, kaons, nucleons and lambdas (which carry both baryonic and strange charge), and which interacts through a larger number of resonances ($\pi K N \Lambda$). The third one consists of the full SMASH simulation, which includes all particles included by the PDG 2019 recension \cite{Tanabashi:2018oca} up to masses of $\sim 2.3$ GeV. Table \ref{particle_content} presents a summary of the present particles and resonances in the first two gases, and Fig. \ref{chem_comp} an overview of the chemical composition of each gas at various temperatures.

\begin{table}
\begin{center}
 \begin{tabular}{||c | c | c| c||} 
 \hline
 Particle & Mass (GeV) & Width (GeV) & Present in \\ [0.5ex] 
 \hline\hline
 $\pi$ & 0.138 & 0 & All\\ 
 $\eta$ & 0.548 & 0 & $\pi K N \Lambda$\\ 
 $\sigma$ & 0.800 & 0.400 & $\pi K N \Lambda$\\ 
 $\rho$ & 0.776 & 0.149 & $\pi K N$,$\pi K N \Lambda$\\ 
 K & 0.494 & 0 & All\\ 
 K\*(892) & 0.892 & 0.050 & $\pi K N$,$\pi K N \Lambda$\\ 
 N & 0.938 & 0 & All\\ 
 N(1440) & 1.440 & 0.350 & $\pi K N$,$\pi K N \Lambda$\\ 
 N(1520) & 1.515 & 0.110 & $\pi K N \Lambda$\\ 
 N(1535) & 1.530 & 0.150 & $\pi K N \Lambda$\\ 
 N(1650) & 1.650 & 0.125 & $\pi K N \Lambda$\\ 
 $\Delta$ & 1.232 & 0.117 & $\pi K N \Lambda$\\ 
 $\Lambda$ & 1.116 & 0 & $\pi K N \Lambda$\\[1ex] 
 \hline
\end{tabular}
\caption{Properties of particles present in the simpler considered hadron gases: $\pi K N$-hard, $\pi K N$, and $\pi K N \Lambda$. Many more particles are present in the full hadron gas, see \cite{Weil:2016zrk}.}
\label{particle_content}
\end{center}
\end{table}

We follow the methodology described in previous papers on shear viscosity \cite{Rose:2017ntg,Rose:2017bjz} and electric conductivity \cite{Hammelmann:2018ath}, and first calibrate non-physical parameters of the Green-Kubo exponential fitting by comparing the result to a semi-analytical calculation based on the Chapman-Enskog formalism\footnote{Note that the cross-conductivities and the diffusion coefficient matrix from \cite{Greif:2017byw,Fotakis:2019nbq} are strongly related via: $\sigma_{Qq} = \kappa_{Qq}/T$, where $q$ is an arbitrary conserved charge and $Q$ is the electric charge.} \cite{Hammelmann:2018ath,Greif:2016skc,Greif:2017byw,Fotakis:2019nbq}. Specifically, as shown on Fig. \ref{calibration_cross}, we calculate the electric, baryonic-electric and strange-electric conductivities for the previously introduced $\pi K N$-hard hadronic gas (with a constant cross-section of 30 mb). We see that there is agreement between the calculations at all temperatures within statistical error bars for the case of the electric and strange-electric conductivities. The baryonic-electric conductivity shows more variation, especially at low temperatures, since there is only a smaller number of protons present leading to larger potential systematic errors (especially numerical as the fluctuations of the correlation function are much smaller than for the other two cases). We use this calibration calculation to establish this systematic error to 10\% for the electric conductivity, 20\% for the baryonic-electric conductivity and 15\% for the strange-electric conductivity. Statistical errors are smaller than the symbol size, and are thus neglected for the remainder of this work. All errors on further figures stem from this estimation of the systematic error.

\begin{figure}
  \centering
  \includegraphics[width=76mm]{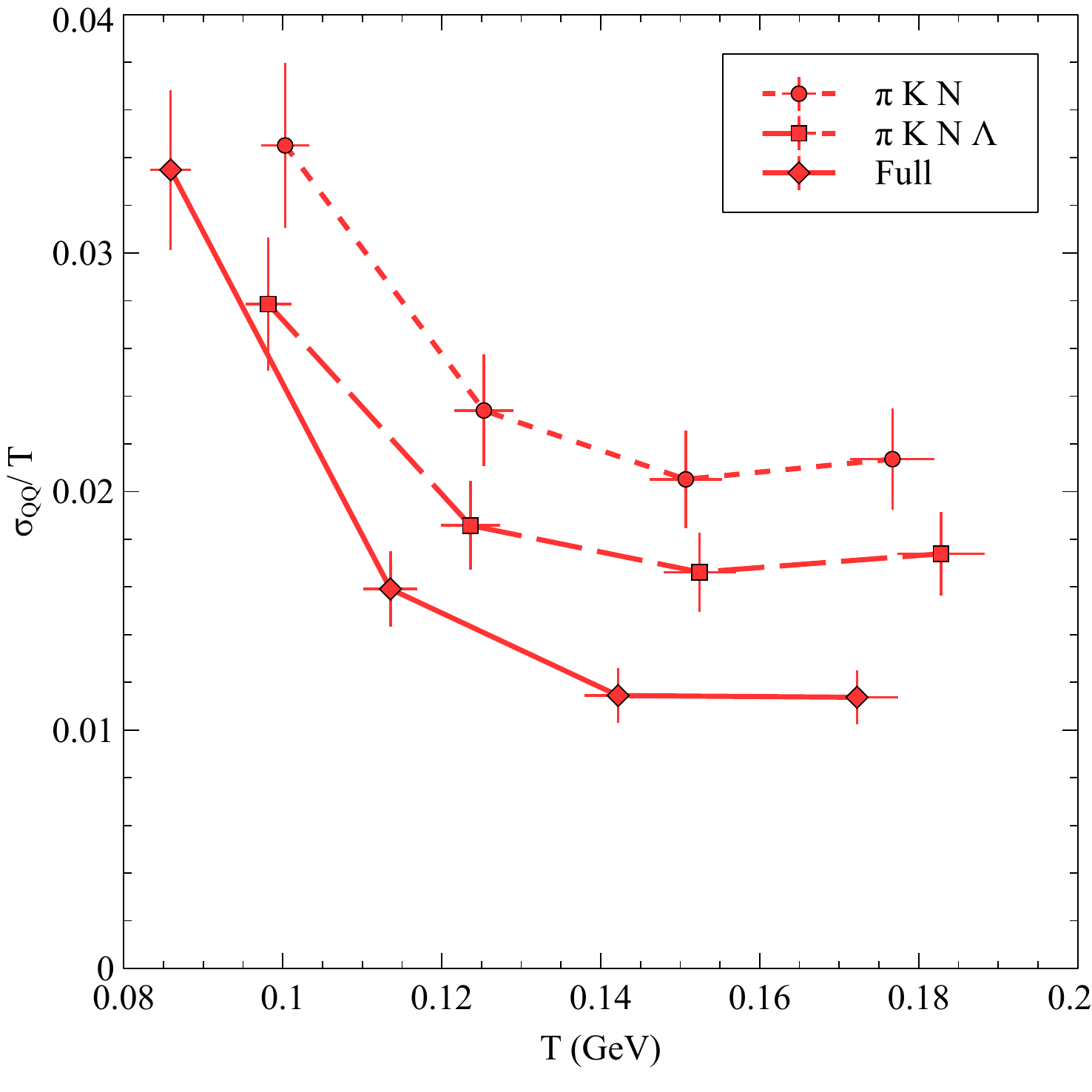}
  \caption{Electric conductivity for resonant hadron gases of increasing complexity.}
  \label{comparative_QQ}
\end{figure}

Let us now discuss the influence of varying the number of active degrees of freedom by looking at the above introduced systems with different chemical composition. In Fig.~\ref{comparative_QQ} the electric conductivity is shown for all three gases. All curves show a decreasing behavior as temperature increases, which eventually levels into a plateau around $T=150$ MeV (and possibly a slight increase onwards for simpler gases; this is consistent with previous calculations \cite{Hammelmann:2018ath,Greif:2016skc}). We observe that as the number of degrees of freedom increases in the system, the general observed behavior is a decrease at every temperature, with this decrease being more pronounced at larger temperatures. This is expected, as adding more resonances into the gas mostly amounts to increasing the cross-sections of the stable particles\footnote{The increased number of degrees of freedom also increases the total density of the system and thus the scattering rates of all particles. It was argued in \cite{Fotakis:2019nbq} that all diffusion coefficients scale inversely with the scattering rate, which then adds a sub-leading decreasing contribution to the temperature profile of the coefficients.} (i.e. pions, kaons, nucleons and eventually heavier strange particles such as the $\Lambda$). For example, at low temperatures, there is significant decrease when going from the $\pi K N$ to the $\pi K N \Lambda$ system, but not a large difference between the $\pi K N \Lambda$ and the full hadron gas. This can be explained by remembering that at such low temperatures, the vast majority of particles are pions (see Fig. \ref{chem_comp}), and as such the inclusion of the $\sigma$ resonance, which is the second largest contribution to the total $\pi \pi$ cross-section after the $\rho$, in the $\pi K N \Lambda$ gas makes a large difference to the total electric conductivity. Adding all the other resonances contributing to the $\pi \pi$ cross-section in the full hadron gas still makes a small difference, but its magnitude is much less, since there is simply not enough energy density in the gas to produce these heavier resonances at this low temperature.

\begin{figure}
  \centering
  \includegraphics[width=76mm]{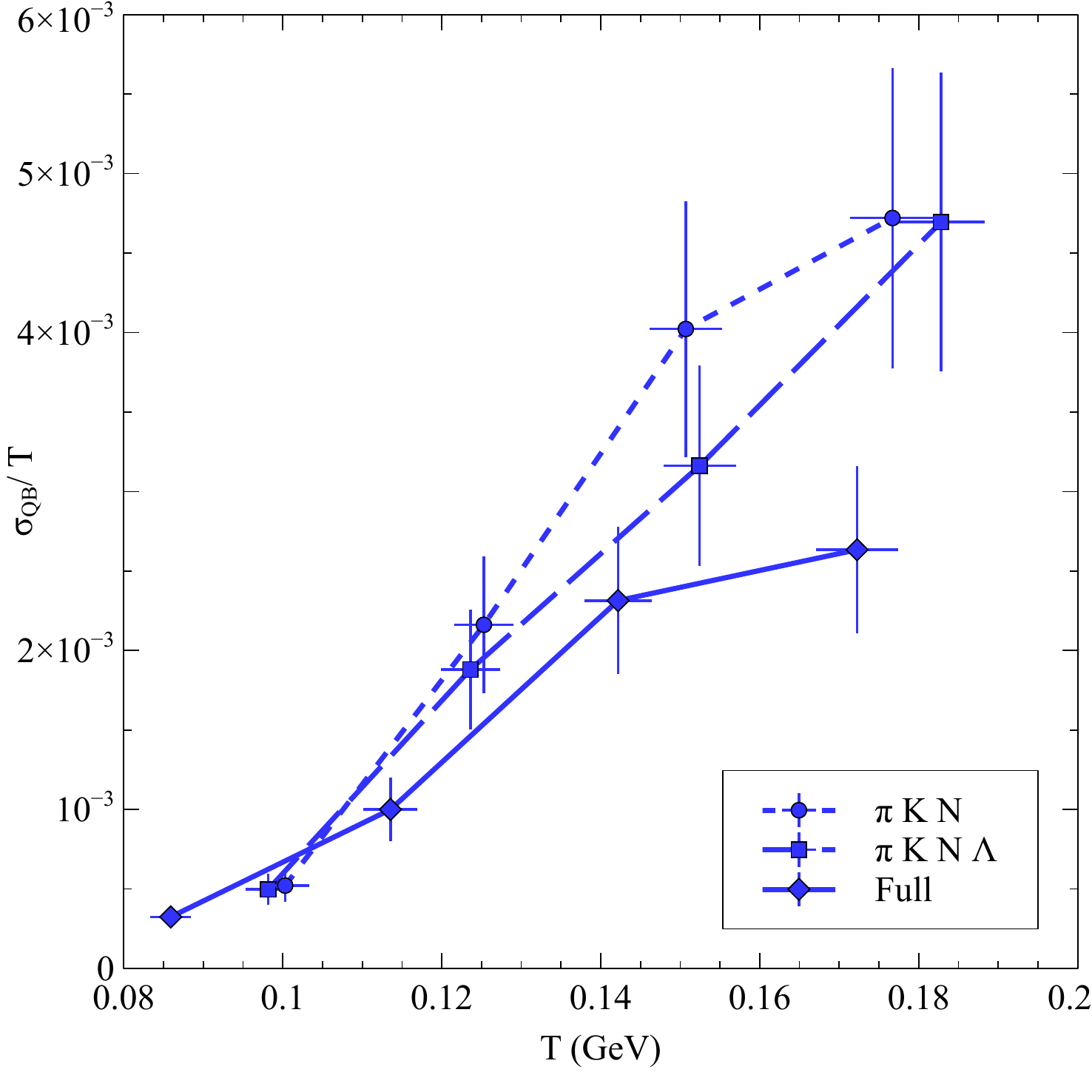}
  \caption{Baryonic-electric conductivity for resonant hadron gases of increasing complexity.}
  \label{comparative_QB}
\end{figure}

Fig.~\ref{comparative_QB} shows the baryonic-electric conductivity as a function of temperature for the same increasingly complex gases. The general trend is that this conductivity increases with temperature, which is expected as the proportion of baryons in a gas should increase with temperature, since they have a comparatively higher mass than mesons (this is something we do observe in Fig. \ref{chem_comp}). Although large uncertainties blur the picture up to 140 MeV, we once again see an ordering in the three cases, with the baryonic-electric conductivity decreasing as the complexity increases, since, as mentioned previously, one of the main effects of increasing the number of degrees of freedom is to increase the cross-sections of the abundant stable particles. We observe a clear separation between the full hadron gas case and the other two simpler cases at temperatures above 160 MeV, with the full hadron gas being markedly lower, which can be explained by the fact that at these temperatures many heavy baryonic resonances which were not included in the simpler gases become relevant.

\begin{figure}
  \centering
  \includegraphics[width=76mm]{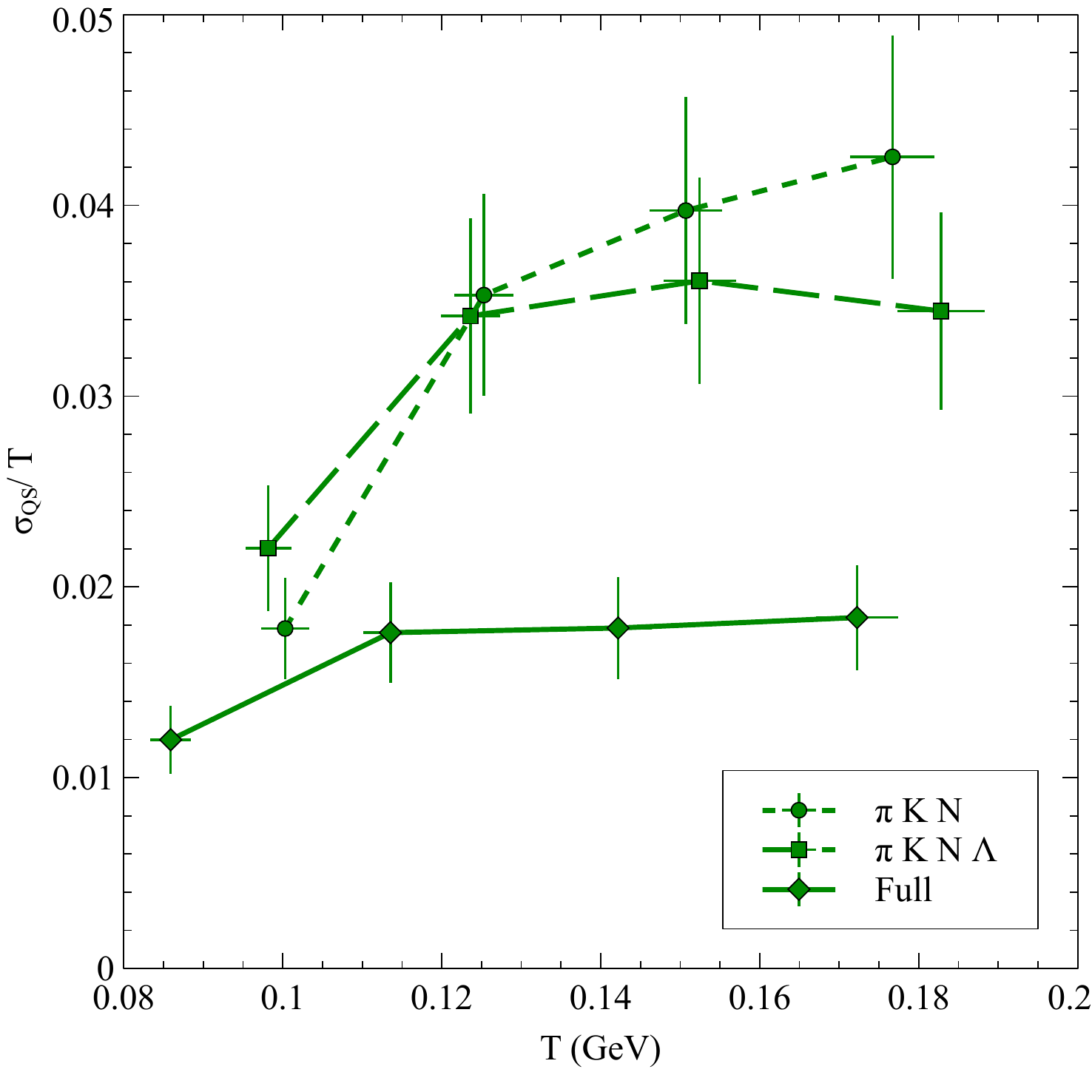}
  \caption{Strange-electric conductivity for resonant hadron gases of increasing complexity.}
  \label{comparative_QS}
\end{figure}

Lastly, the strange-electric conductivity as a function of temperature is presented in Fig.~\ref{comparative_QS}. All three curves exhibit a rapidly increasing behavior at low temperature, which then appears to slow down as temperature increases, with all three gases also being fully consistent within uncertainties with reaching a plateau around 120 MeV. This behavior can be understood by looking at the chemical composition of the gases: at low temperature, the proportion of the light strangely charged kaon increases very fast with respect to other types of particles, but eventually slows down because of the inclusion of more and more non-strange heavier particles. A plateau is reached at larger temperatures when similar numbers of strange and non-strange particles are added into the system. This coefficient exhibits a much larger response to adding large amounts of states into the mixture as the previous ones, as the two simpler cases remain very close to each other at all but the very highest temperatures. We note that this difference between the $\pi K N$ and $\pi K N \Lambda$ cases is most likely due to the introduction of a fourth strange stable particle, the $\Lambda$, which becomes more relevant at these high temperatures and could explain the small dip in the value of the strange-electric conductivity. The full hadron gas' conductivity remains lower, which could be very helpful to discriminate between sufficient and insufficient amounts of states in a given hadron gas model, if the exact value of this transport coefficient can be confirmed for example with lattice QCD calculations.

In summary, we introduced novel conductivity transport coefficients that quantify the impact of the electrical field on the generation of currents of various conserved quantum numbers, e.g. baryon number and strangeness. We have presented the first calculation of the baryon-electric, strange-electric and purely electric conductivity as a function of temperature in a hadron gas with varying degrees of freedom. A simple gas containing pions, protons and kaons was used to compare to Chapman-Enskog calculations and determine the systematic uncertainty of our method. For all 3 transport coefficients a large sensitivity to the active degrees of freedom in the hadron gas is found. Thus, we conclude that the chemical composition of the gas is very relevant to the coupling of the electric field with the system and the corresponding generation of currents for all considered conserved quantum numbers. If future lattice QCD calculations provide firm values for the cross-conductivities, they can be used to determine the composition of the hadron gas, potentially identify missing states and estimate the strength of hadronic interactions. In addition, the results provided here are relevant as an input to hydrodynamic simulations that include multiple conserved charges explicitly.

\begin{acknowledgments}

The authors would like to acknowledge J.M. Torres-Rincon for his contribution to this work through fruitful discussions.
J.-B. Rose's work was supported by the Helmholtz International
Center for the Facility for Antiproton and Ion Research (HIC for FAIR) within the
framework of the Landes-Offensive zur Entwicklung Wissenschaftlich-Oekonomischer Exzellenz (LOEWE) program launched by the State of Hesse. Computational resources have been provided by the Center
for Scientific Computing (CSC) at the Goethe-University of Frankfurt. This work was supported by the Deutsche Forschungsgemeinschaft (DFG, German Research Foundation) – Project number 315477589 – TRR 211.
M.G. and J.A.F.
    acknowledge support from the \textquotedblleft Helmholtz Graduate School
    for Heavy Ion research\textquotedblright. J.A.F. acknowledges support from the  \textquotedblleft Stiftung Polytechnische
    Gesellschaft\textquotedblright, Frankfurt am Main.  Furthermore, G.S.D. would like to thank Conselho Nacional de
    Desenvolvimento Cient\'{\i}fico e Tecnol\'{o}gico (CNPq) and Funda\c c\~ao de Amparo \`a Pesquisa do Estado do Rio de Janeiro (FAPERJ) for financial support.
\end{acknowledgments}

\end{document}